\documentclass[aps,twocolumn,amsmath,amssymb,reprint,numbers,superscriptaddress,noeprint,longbibliography]{revtex4-1}

\usepackage{braket}
\usepackage{soul}
\usepackage{geometry}
\usepackage{graphicx}
\usepackage{booktabs}
\usepackage{array}
\usepackage{relsize}
\usepackage{color}
\usepackage[dvipsnames]{xcolor}
\usepackage{xcolor}
\usepackage{eqnarray}
\usepackage{textcase}
\usepackage{textcomp}
\usepackage{setspace}
\usepackage{siunitx}
\usepackage{placeins}
\usepackage{textgreek}
\usepackage{xcolor}
\usepackage{tikz}
\usetikzlibrary{positioning}
\usepackage{hyperref}
\hypersetup{
     colorlinks   = true,
     linkcolor    = blue,
     citecolor    = blue,
     urlcolor = blue
}

\definecolor{cola}{rgb}{0.7,0.1,0.1}
\definecolor{colb}{rgb}{0.9,0.4,0}
\definecolor{colc}{rgb}{0.3,0.7,0}
\definecolor{cold}{rgb}{0,0.35,0.75}
\definecolor{cole}{rgb}{0.63, 0.13, 0.94}
\definecolor{colf}{rgb}{0.5, 0.5, 0.5}
\definecolor{colred}{rgb}{1, 0, 0}

\newcommand{\Tone}{$T_1$}

\newcommand{\Do}{D\textsuperscript{0}}
\newcommand{\DoX}{D\textsuperscript{0}X}

\newcommand{\AlZn}{Al\textsubscript{Zn}}
\newcommand{\GaZn}{Ga\textsubscript{Zn}}
\newcommand{\InZn}{In\textsubscript{Zn}}

\newcommand{\AlZnN}{Al$_\mathrm{Zn}^\mathrm{0}$}
\newcommand{\GaZnN}{Ga$_\mathrm{Zn}^\mathrm{0}$}
\newcommand{\InZnN}{In$_\mathrm{Zn}^\mathrm{0}$}
\newcommand{\InP}{In\textsuperscript{+}}
\newcommand{\InZnP}{In$_\mathrm{Zn}^\mathrm{+}$}

\definecolor{col2}{rgb}{0,1,0}

\begin{document}

\title[]{Properties of Donor Qubits in ZnO Formed by Indium-Ion Implantation}

\author{Xingyi Wang}\thanks{The authors contribute equally to this project. \\Corresponding author: xingyiw@uw.edu}
\affiliation{Department of Electrical Engineering, University of Washington, Seattle, Washington 98195, USA}
\author{Christian Zimmermann}\thanks{The authors contribute equally to this project. \\Corresponding author: xingyiw@uw.edu}
\affiliation{Department of Physics, University of Washington, Seattle, Washington 98195, USA}
\author{Michael Titze}
\affiliation{Sandia National Laboratories, Albuquerque,
New Mexico 87123, United States}
\author{Vasileios Niaouris}
\affiliation{Department of Physics, University of Washington, Seattle, Washington 98195, USA}
\author{Ethan R. Hansen}
\affiliation{Department of Physics, University of Washington, Seattle, Washington 98195, USA}
\author{Samuel H. D'Ambrosia}
\affiliation{Department of Physics, University of Washington, Seattle, Washington 98195, USA}
\author{Lasse Vines}
\affiliation{Department of Physics/Centre for Materials Science and Nanotechnology, University of Oslo, Blindern, Oslo N-0316, Norway}
\author{Edward S. Bielejec}
\affiliation{Sandia National Laboratories, Albuquerque,
New Mexico 87123, United States}
\author{Kai-Mei C. Fu}
\affiliation{Department of Physics, University of Washington, Seattle, Washington 98195, USA}
\affiliation{Department of Electrical Engineering, University of Washington, Seattle, Washington 98195, USA}
\affiliation{Physical Sciences Division, Pacific Northwest National Laboratory, Richland, Washington 99352, USA}

\begin{abstract}
Neutral shallow neutral donors (D$^\mathrm{0}$) in ZnO have emerged as a promising candidate for solid-state spin qubits. 
Here, we report on the formation of D$^\mathrm{0}$ in ZnO via implantation of In and subsequent annealing. 
The implanted In donors exhibit optical and spin properties on par with those of \emph{in situ} doped donors.
The inhomogeneous linewidth of the donor-bound-exciton transition is less than \SI{10}{\giga\hertz}, comparable to the optical linewidth of \emph{in situ} In. 
Longitudinal spin relaxation times ($T_1$) exceed reported values for \emph{in situ} Ga donors, indicating that residual In-implantation damage does not degrade $T_1$. 
Two-laser Raman spectroscopy of the donor spin reveals the hyperfine interaction of the donor electron with the spin-9/2 In nuclei. 
This work is an important step toward the deterministic formation of In-donor qubits in ZnO with optical access to a long-lived nuclear-spin memory.
\end{abstract}

\date{\today}

\maketitle

\section{Introduction}
Neutral shallow donors (\Do) in semiconductors have shown potential as solid-state spin qubits for use in quantum technologies, such as quantum computing and quantum communication\,\cite{MortonQubits, YamamotoQubits, VanderspyenQubits, HeQubits}. 
In direct-band-gap semiconductors, the bound electron spin states of shallow donors forming the qubit states can be optically accessed via the donor-bound exciton (\DoX) with high radiative efficiency\,\cite{XiayuDirectT1, ClarkD0X}.
The direct-band-gap semiconductor ZnO is an emerging platform for  \Do\ qubits, with D typically consisting of Al, Ga, or In substituted on a Zn site, denoted as \AlZn, \GaZn, or \InZn, respectively. 
For these donor qubits, narrow optical \DoX\ linewidths, efficient radiative transitions, optical state initialization, long longitudinal spin relaxation times ($T_1>400$\;ms) and moderate coherence times ($T_2\approx 50$\;ms) have been demonstrated\,\cite{VasilisT1, XiayuZnO, MariaZnO}.

One feature of qubits based on impurities is that impurities can be incorporated by ion implantation and subsequent annealing\,\cite{TitzeIonImpl, ChakravarthiIonImpl, MccallumIonImpl, JakobIonImpl}. 
With use of ion implantation, donor density and depth can be controlled. 
Thus, one can introduce either high densities of donors for quantum memory and transduction applications~\cite{LvovskyQM, williamson2014mom} or low densities of individually addressable donor qubits for quantum computing and networks\,\cite{YamamotoQubits, MortonQubits, LilieholmYb}. 
Using focused ion beam technology, one can also control the lateral positioning of the donor impurities\,\cite{PachecoFIB, TamuraFIB}, enabling the deterministic placement of single donor qubits after fabrication of photonic devices~\cite{schroder2017SFI}. 
Moreover, for single site defects, such as substitutional donors, near-deterministic incorporation is possible~\cite{JakobIonImpl}.

Here, we report on the optical and spin properties of \InZnN\ in ZnO formed at a depth of \SI{200}{\nano\metre} through ion implantation and subsequent annealing.
We focus on \InZnN\ because of the following favorable features: large binding energy ($\approx$~\SI{63.2}{\milli\electronvolt}\,\cite{WagnerD0X}), strong hyperfine interaction with the In nucleus\,\cite{BlockInODMR, GonzalezInODMR} for access to a nuclear spin quantum memory \cite{MortonQM}, and the availability of ZnO substrates with low residual In doping. 
Ion implantation at relatively large fluences has previously been employed to yield impurity-related shallow donors in ZnO\,\cite{MoholtSn, CullenIn, CullenGe, MullerIn, ZatsepinImpl}. 
Here we focus on low implantation fluences and the resulting donor properties relevant to quantum information applications. 
We demonstrate that implanted \InZnN\ exhibit \DoX\ inhomogeneous optical transition linewidths less than \SI{10}{\giga\hertz}, comparable to the narrowest linewidths observed for in-grown \InZnN. 
The narrow linewidths owing to low strain and small spectral diffusion from charge traps indicate low implantation damage, which is consistent with the reported resilience of ZnO to radiation damage\,\cite{KucheyevImpl, KucheyevImpl2}.
Favorable spin properties are also demonstrated, with measured \Tone\ times in the hundreds of milliseconds, approximately 4 times longer than the \Tone\ times observed for in-grown \GaZnN\ donors\,\cite{VasilisT1}. 
Further, two-laser Raman spectroscopy on implantation-doped \InZnN\ demonstrates that we can begin to resolve the large hyperfine interaction between the bound electron and the In nucleus, pointing to the potential for an optical interface to a long-lived nuclear spin memory.

\section{Sample Preparation}
\begin{figure}[!htb]
  \centering
  \includegraphics[width=0.43\textwidth]{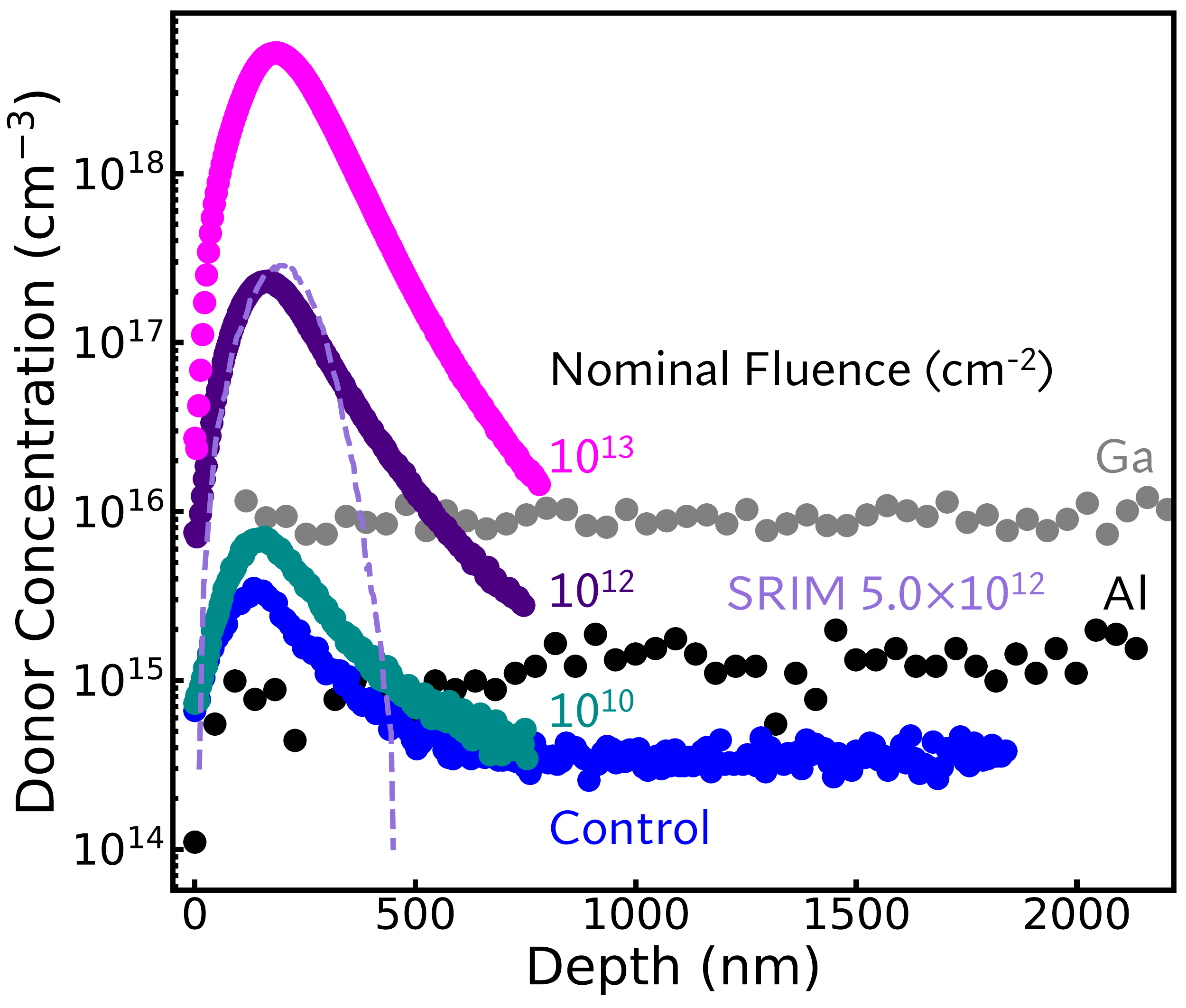}
  \caption{SIMS data collected from the central 62\;\textmu m of a crater size of 200\;\textmu m: The measured (nominal) In implantation fluences are 1.1$\times$10$^{14}$ (10$^{13}$ cm$^{-2}$)\,cm$^{-2}$, 5.0$\times$10$^{12}$ (10$^{12}$ cm$^{-2}$)\,cm$^{-2}$, 1.7$\times$10$^{11}$ (10$^{10}$ cm$^{-2}$)\,cm$^{-2}$,
  1.3$\times$10$^{11}$ (Control)\,cm$^{-2}$. The peak of In concentration from the SRIM simulated In implantation profile matches well with that of the measured implantation profile.}
  \label{fig:sims} 
\end{figure}

A hydrothermally-grown ZnO single crystal substrate (Tokyo Denpa, \SI{5}{\milli\metre} $\times$\,\SI{5}{\milli\metre}\,$\times$\,\SI{0.3}{\milli\metre}) with a surface orientation of (0001) is used as the implantation substrate~\cite{MaedaTokyoDenpa} (see Appendix A). 
In the remainder of the article, the [0001] direction of the crystal is referred to as the $\vec{c}$ axis. 
Ion implantation is performed with the Sandia Ion Beam Laboratory \SI{6}{\mega\volt} Tandem accelerator. 
In is extracted as a negative ion utilizing a pressed cathode filled with In wire. 
The landing energy of \InP\ ions is \SI{800}{\kilo\electronvolt}, leading to an implant depth of approximately \SI{200}{\nano\metre} determined by our simulating the implantation process using the Stopping and Range of Ions in Matter software\,\cite{ZieglerSRIM} (Fig.~\ref{fig:sims}). 
The ions are focused with a magnetic quadrupole lens, giving a beam spot of $780 \times 870$ \textmu m$^2$. 
The spot size is measured with a luminescent phosphor and imaged with a home-made camera setup. 
The imaging setup uses a \SI{10}{\milli\metre} hole in the center of its final imaging mirror, allowing concentric imaging of the beam. 
The focused ion beam allows us to vary the ion implantation fluence within a single sample. 
The nominal implantation fluence ranges from 10$^8$ to 10$^{15}$ ions/cm$^2$. Post-implantation annealing at 700 $^{\circ}$C in oxygen for 1 h is used to recover implantation damage. 

Secondary ion mass spectrometry (SIMS) is performed to characterize the In implantation profile and to determine the background concentrations of Al and Ga, prominent donors present in the substrate~\cite{VinesImpurityConcentration, WagnerD0X, MeyerD0X}. 
A Cameca IMS 7f secondary ion mass spectrometer equipped with a 15\;keV $\text{O}_2$ primary ion beam source is used to record the concentration vs. depth profiles of In, Al and Ga. 
Absolute concentrations of In are obtained by our measuring  ion implanted reference samples, ensuring less than $\pm$10\;\% error in accuracy. 
For depth calibration, the sputtered crater depths are determined by a Dektak 8 stylus profilometer and a constant erosion rate is assumed.

As shown in Fig.~\ref{fig:sims}, the measured peak implantation depth is consistent with the depth from simulation. 
However the measured implantation fluence ranges from 5 to more than 10 times the nominal fluence. 
Moreover, a fluence of 10$^{11}$\,cm$^{-2}$ is measured in a nominally unimplanted control region. 
The discrepancy in the nominal and measured implantation fluences is attributed to overspray from the very high fluence implantation areas. 
Despite the higher-than-intended fluences, as we show below, the optical and spin properties of implanted In donors appear promising for quantum information applications. 
Throughout this work we refer to the implantation regions in terms of the nominal implantation fluence. 
Finally, a background concentration of 9.2$\times$10$^{15}$\,cm$^{-3}$ and 1.2$\times$10$^{15}$\,cm$^{-3}$ is measured for Ga and Al respectively.

\section{Results and Discussion}

The properties of the implanted In donors are characterized via low-temperature photoluminescence (PL) and PL excitation (PLE) spectroscopy, spin relaxation and two laser coherent population trapping (CPT) measurements. 

\subsection{Photoluminescence Spectroscopy}
\begin{figure*}[t]
  \centering
  \includegraphics[width=1\textwidth]{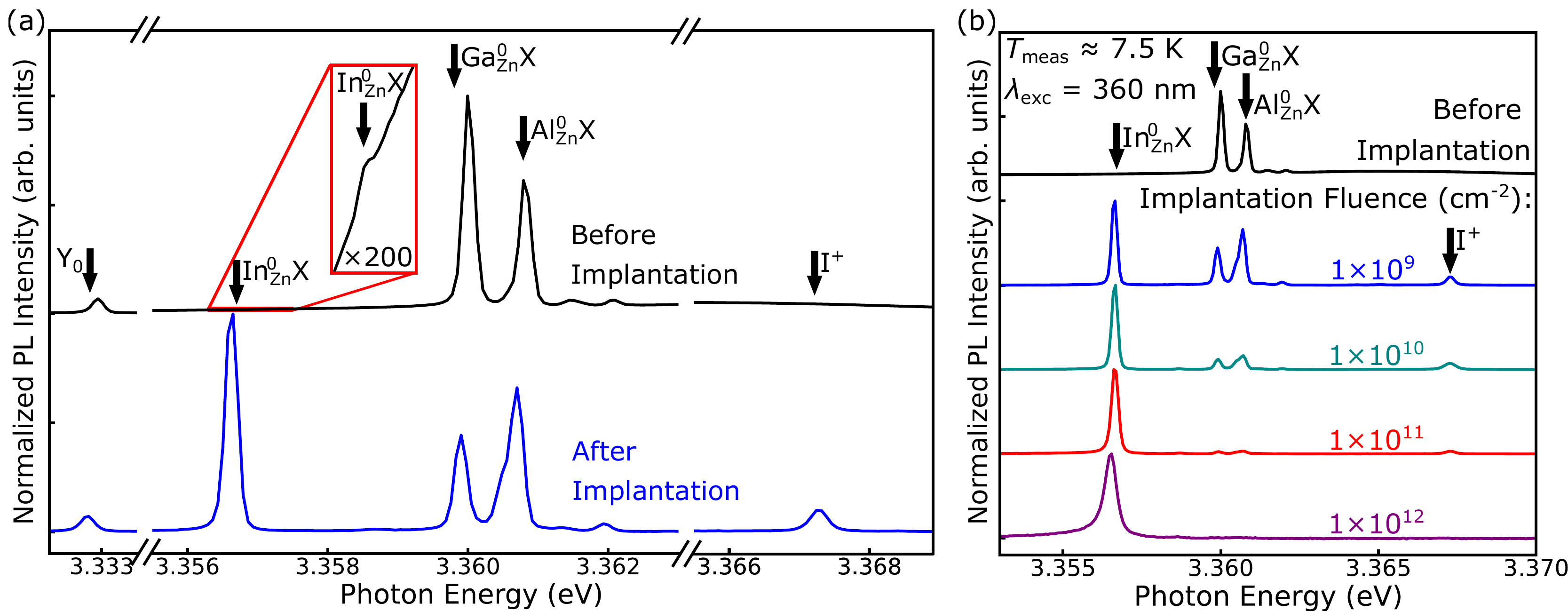}
  \caption{\label{fig:PL_overview} (a) Comparison of normalized PL spectra of an as-grown ZnO single crystal, and for a single crystal after In implantation and subsequent annealing. (b) PL spectra normalized to  the highest-intensity peak for different implantation fluences. $T=7.5$\,K, $\lambda_{\text{exc}} = 360$\,nm, $I_{\text{exc}}=800$\,nW/\textmu m\textsuperscript{2}.}
\end{figure*}

The optical signature of donor incorporation in a semiconductor is the corresponding \DoX\ luminescence transition. 
Fig.~\ref{fig:PL_overview}~(a) shows a comparison of the normalized PL spectra from an as-grown ZnO single crystal and for a single crystal after In implantation and subsequent annealing. 
Before In implantation, a weak signal attributed to the \InZnN X line\cite{WagnerD0X} can be observed. 
The corresponding feature is about three orders of magnitude weaker than the PL features attributed to \GaZnN X and \AlZnN X\,\cite{WagnerD0X}, indicating that \InZnN\ is present in concentrations on the order of 10$^{13}$\;cm$^{-3}$. 
After In implantation, a dramatic increase in the \InZnN X line relative to the \GaZnN X and \AlZnN X lines is observed. 
The \InZnN X linewidth is spectrometer resolution limited up to a fluence of 10$^{11}$\;cm$^{-2}$.
The complete fluence-dependent spectra are included in Appendix B.

In a  ddition to the prominent \InZnN X feature, a few additional new features can be observed after implantation and annealing. 
A sharp line at \SI{3.3673}{\electronvolt} is labelled I$^\mathrm{+}$ (see Fig.~\ref{fig:PL_overview}~(a)). 
This feature is close to a PL feature attributed to excitons bound to ionized In$_\mathrm{Zn}$\,\cite{KumarPLZnO}. 
Magneto-PL measurements on this sample confirm that the I$^\mathrm{+}$ is related to an exciton bound to either an ionized donor or acceptor (see  Appendix C) and we attribute the I$^\mathrm{+}$ line to In$_\mathrm{Zn}^\mathrm{+}$X. 
Additionally, a low-energy shoulder on the \AlZnN X line can be observed, indicating additional donor formation of unknown origin (see  Appendix C). 
Finally, the prominent Y$_\mathrm{0}$ exciton line, which is thought to arise from excitons bound to structural defects\,\cite{WagnerD0X}, can be seen in the sample both before and after In implantation.

\begin{figure}[htb!]
  \centering
  \includegraphics[width=1\linewidth]{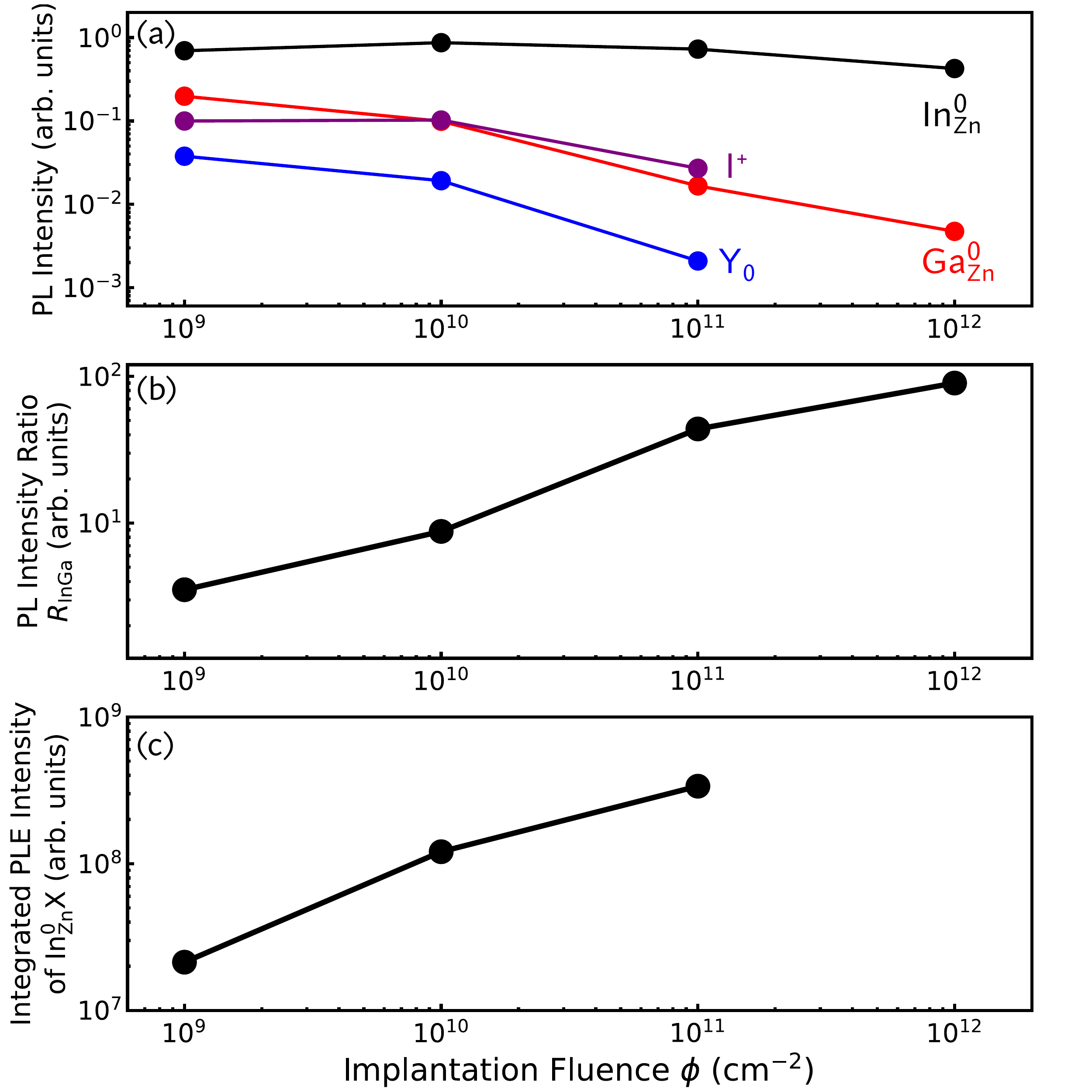}
  \caption{\label{fig:incorporation}
  (a) Dependence of PL intensity on In implantation fluence for transitions labelled in Fig.~\ref{fig:PL_overview}(a). 
  (b) Dependence of the ratio of the integrated PL intensity for \InZnN X to \GaZnN X, denoted $R_\mathrm{InGa}$, on In implantation fluence. 
  (c) Integrated PLE intensity of \InZnN X. 
  $T=7.5$\,K, $\lambda_{\text{exc}} = 360$\,nm (for PL), $I_{\text{exc}}=800$\,nW/\textmu m\textsuperscript{2}.
  }
\end{figure}

Curiously, the absolute PL intensity for \InZnN X remains approximately constant with implantation fluence (Fig.~\ref{fig:incorporation}(a)), whereas the PL intensity of the \AlZnN X, \GaZnN X, and Y$_\mathrm{0}$ decreases with increasing In implantation fluence. 
These observed changes in the integrated PL intensity indicate that the entire volume of material probed with near-band gap excitation ($\lambda_\mathrm{exc}=360$\;nm) is affected by the In implantation. 
This does not necessarily mean the probing depth is only the \SI{200}{\nano\metre} depth of implanted In ions, because intrinsic defects created during implantation could diffuse deeper into the material during annealing\,\cite{AzarovZnDiffusion, SkyInDiffusion, SteiaufInDiffusion}. 

While the absolute \InZnN X remains constant, the ratio of the integrated PL intensity of \InZnN X and \GaZnN X, denoted as $R_\mathrm{InGa}$, increases with fluence, as shown in Fig.~\ref{fig:incorporation}(b). 
$R_\mathrm{InGa}$ cannot be tracked for fluences higher than 10$^{12}$\,cm$^{-2}$ because the \GaZnN X line cannot be resolved at such high fluences. 
The constant \InZnN X PL intensity coupled with the increase in $R_\mathrm{InGa}$ with fluence suggests an implantation-induced decrease in the bound exciton excitation efficiency and/or decrease in radiative efficiency for all donors in the excitation volume. 
Resonant photoluminescence excitation spectroscopy, described further below, supports the hypothesis that it is the off-resonant excitation efficiency and not the radiative efficiency that is primarily affected by implantation fluence.

\subsection{Photoluminescence Excitation Spectroscopy}

\begin{figure}[t]
  \centering
  \includegraphics[width=1\linewidth]{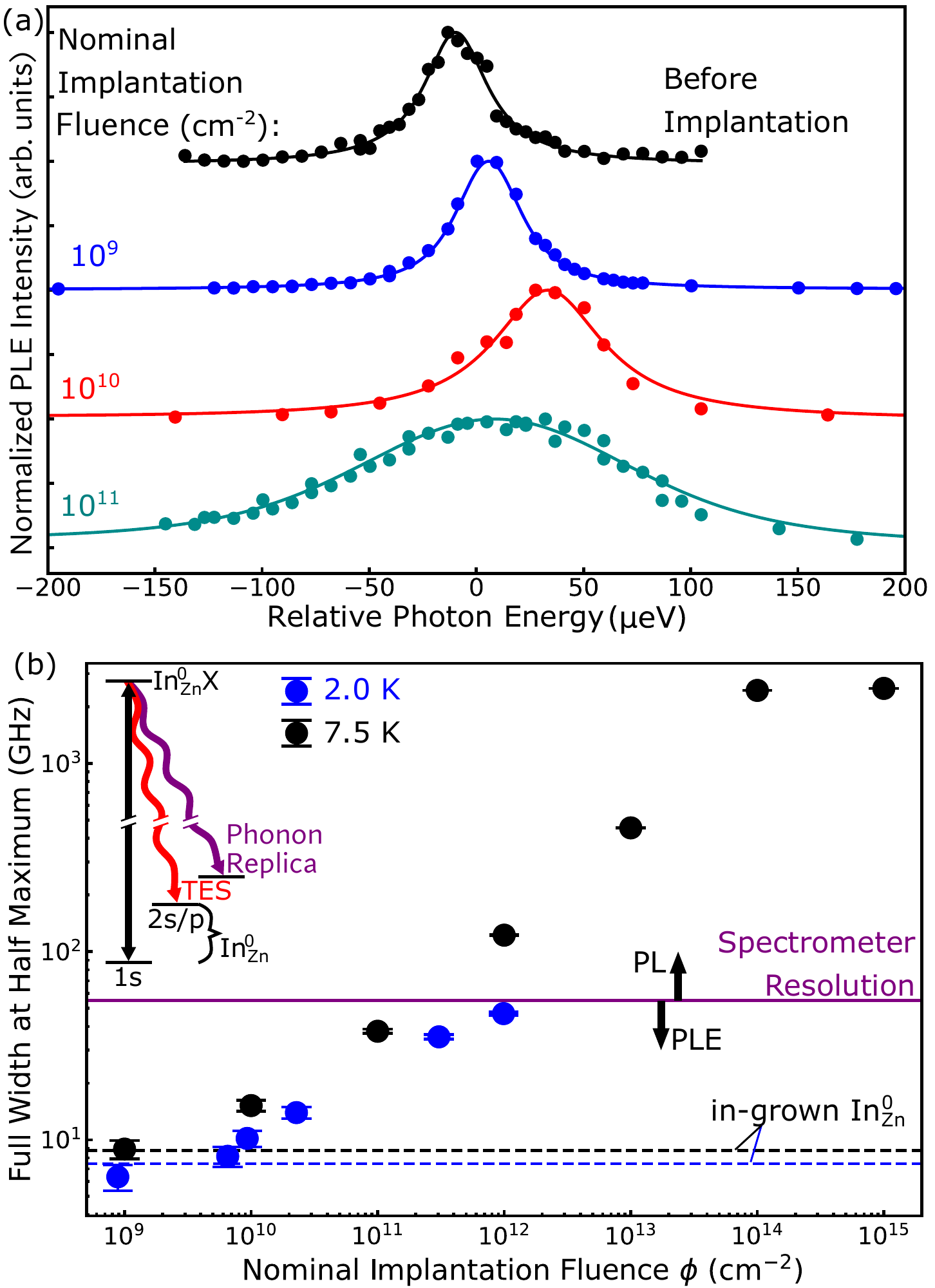}
  \caption{\label{fig:linewidth}(a) PLE spectra of \InZnN X for in-grown and implanted \InZnN. FWHM of each spectrum is determined by our fitting the spectrum to a Voigt profile. 
  (b) Dependence of \InZnN X FWHM on In implantation fluence. The implanted fluence is estimated from $R_\mathrm{InGa}$ for the data at \SI{2.0}{\kelvin}\,(blue dots). 
  Dashed lines indicate the FWHM of in-grown \InZnN X at \SI{2.0}{\kelvin} and \SI{7.5}{\kelvin} performed on a ZnO single crystal with an \InZnN  concentration similar to that of pre-implanted sample.}
\end{figure}

In photoluminescence excitation (PLE) spectroscopy, a narrow band laser scans the dominant transition, corresponding to relaxation from \InZnN X to the 1s donor state of \InZnN, while two-electron satellites (TES, corresponding to relaxation to donor 2s/2p states, see inset in Fig.\ref{fig:linewidth}(b)) and phonon replica transitions are detected~\cite{MariaZnO}. 
Resonant excitation spectra of the two-electron satellites and phonon replica are shown in Appendix E. 
Fig.~\ref{fig:linewidth}(a) shows the PLE spectra for an as-grown ZnO single crystal and for three different In implantation fluences. 
The ability to perform resonant excitation of the \InZnN X transition (see Fig.~\ref{fig:PL_overview}~(b)) shows that an appreciable population of \InZn\ exists in the desired neutral charge state.

Fig.~\ref{fig:linewidth}(b) displays the full width at half maximum (FWHM) extracted from PLE and PL spectra for all implantation fluences. 
For implantation fluences lower than 10$^{10}$\;cm$^{-2}$, the linewidth of implanted \InZnN X is comparable to that of in-grown \InZnN X. 
For larger fluences, the linewidth of implanted \InZnN X ensembles increases with increasing fluence, indicating that residual lattice damage contributes significantly to the inhomogeneous broadening.
In contrast to PL measurements with 360\,nm excitation, under resonant excitation we observe an increase in the \InZnN X PLE-intensity with implantation fluence (Fig.~\ref{fig:incorporation}~(c)). 
This result indicates that the constant PL intensity with fluence observed in Fig.~\ref{fig:incorporation}(a) is at least in part due to reduced excitation efficiency of D$^0$X. 
We also observe a linear relationship between PLE intensity with the PL ratio $R_\mathrm{InGa}$ at low implantation fluence (see  Appendix D). 
We thus utilize $R_\mathrm{InGa}$ as a proxy for fluence in the tails of the implantation regions for finer fluence sampling (blue data in Fig.~\ref{fig:linewidth}(b)).

\begin{figure}[t]
  \centering
  \includegraphics[width=1\linewidth]{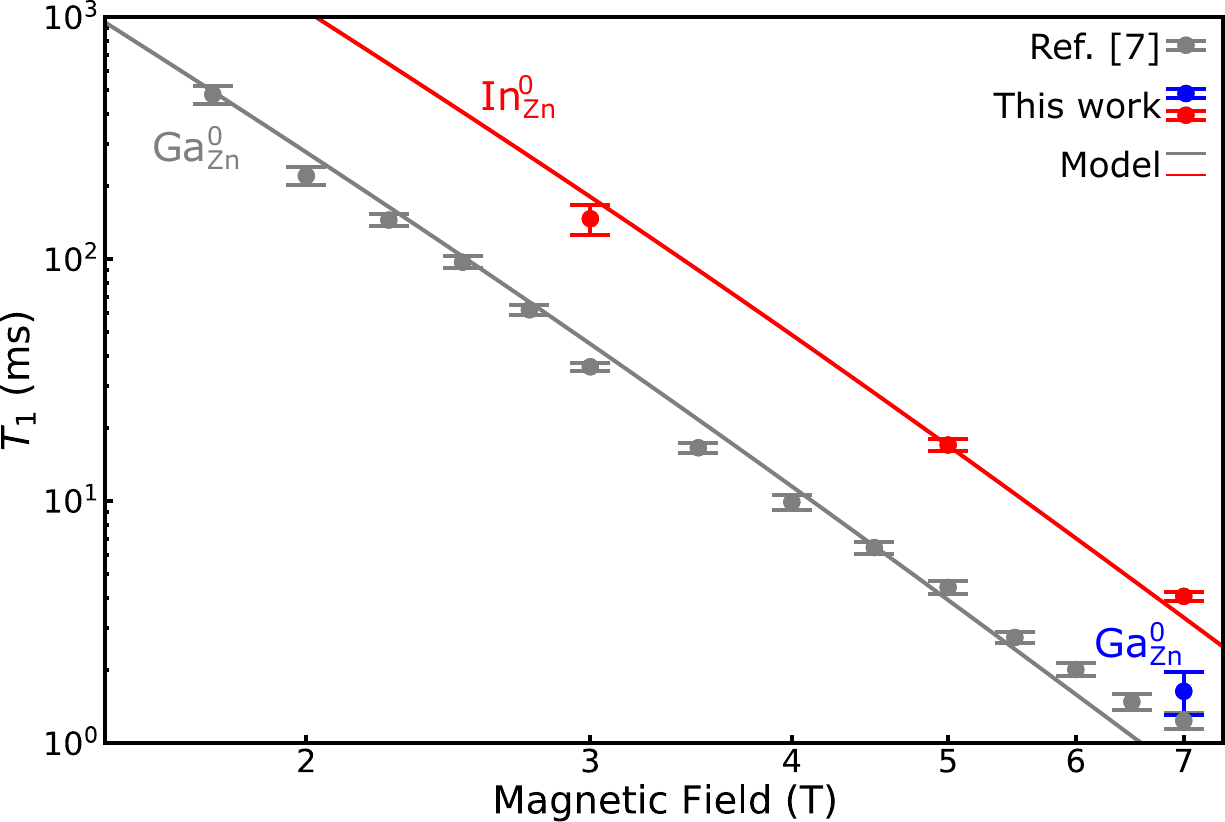}
  \caption{\label{fig:T1}
  Dependence of $T_\mathrm{1}$ on the applied magnetic field. The population is optically initialized into the $\ket{\uparrow}$ state. Relaxation to the thermal equilibrium is probed optically (see Appendix F).  
   Faraday geometry ($\vec{B}\parallel \vec{c}$), $T=1.5$\;K (for~\cite{VasilisT1}), $1.9$\;K (this work). }
  Red: Implanted In, fluence 10$^{9}$cm$^{-2}$. Blue: In-grown Ga. Grey: In-grown Ga from a similar ZnO crystal, published in Ref.~\cite{VasilisT1}. 
\end{figure}

\begin{figure*}[!t]
  \centering
  \includegraphics[width=1\textwidth]{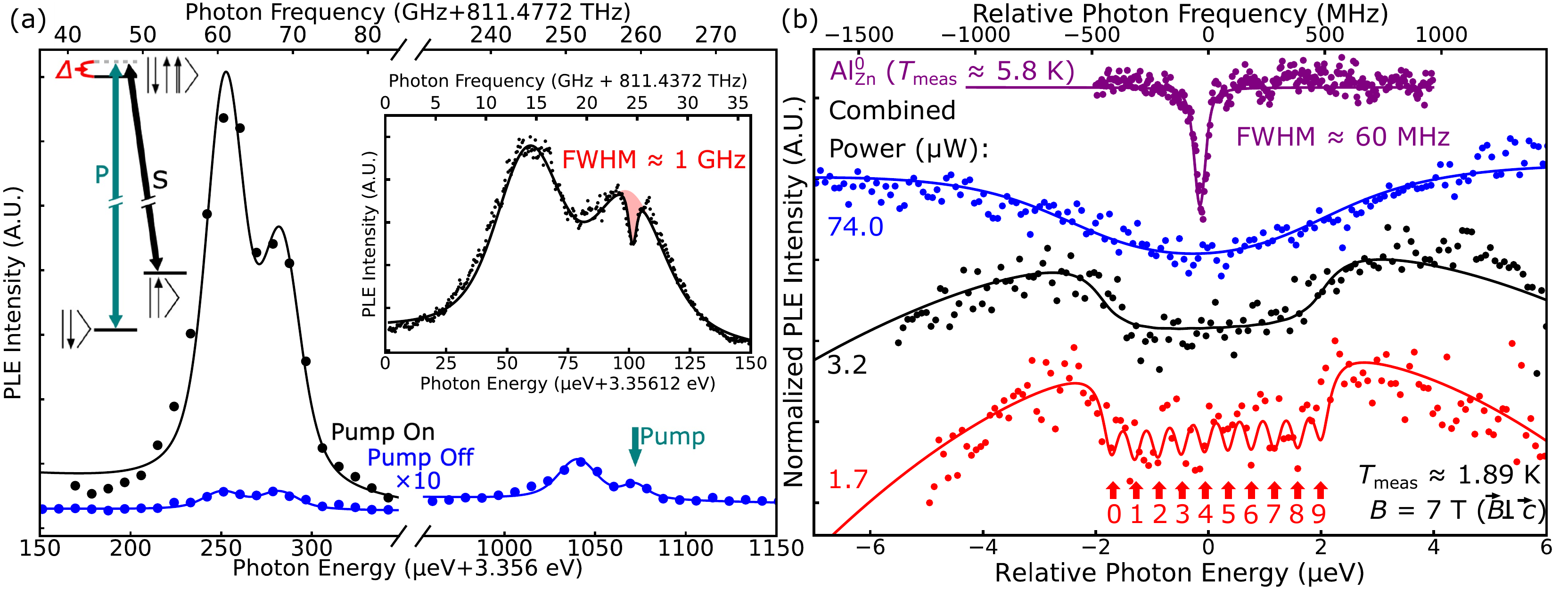}
  \caption{\label{fig:OPnCPT}
  (a) Resonant one-laser and two-laser spectroscopy on implanted (10$^{9}$\;cm$^{-2}$) \InZnN. For the one-laser measurement (blue), a tunable laser scans all four transitions.
  For the two-laser measurement (see inset on the left for the energy diagram, where the dashed gray level denotes the detuning of both lasers from the state $\ket{\downarrow\uparrow\Uparrow}$), the pump laser (P) is vertically polarized and resonant with the $\ket{\downarrow}$ to $\ket{\downarrow\uparrow\Uparrow}$ transition, while the scanning laser (S) is horizontally polarized and tuned across the two transitions involving the $\ket{\uparrow}$ state. 
  A high-resolution scan of the two-laser experiment is depicted in the right inset, in which a CPT dip is observed. 
  (b) High-resolution PLE spectra showing CPT for different combined powers of the two lasers.
  For a constant wavelength of the pump laser, this leads to ten different resonance conditions depending on the nuclear spin m$_\mathrm{I}$. 
  The data are fitted to the sum of ten equally-spaced Lorentzian profiles (splitting equal to $A\approx 100$\;MHz). 
  The full electron-nuclei energy diagram can be found in Appendix H. 
  The measurements are performed at \SI{7}{\tesla} (Voigt geometry $\vec{B} \perp \vec{c}$) and \SI{2}{\kelvin}. 
  }
\end{figure*}

\subsection{Longitudinal Spin Relaxation}

For in-grown \GaZnN, we have previously shown that the dominant mechanism for spin relaxation for the donor bound spin-1/2 electron is due to spin-orbit coupling and the piezoelectric electron-phonon interaction~\cite{VasilisT1}. Here, we perform similar longitudinal spin relaxation measurements for implanted \InZnN. 
Fig.~\ref{fig:T1} shows the dependence of \Tone\ on $B$ in the Faraday geometry at \SI{1.9}{\kelvin} for in-grown \GaZnN\ and implanted \InZnN. 
The dependence of \Tone\ on $B$ can be described by
\begin{equation*}
  \label{eq:T1}
  T_\mathrm{1} =
  \left(\Gamma_{\downarrow\uparrow}+\Gamma_{\uparrow\downarrow}\right)^{-1}
  = \frac{1}{\Gamma_{\downarrow\uparrow}}\frac{\exp\left(\gamma\right) - 1}{\exp\left(\gamma\right) + 1},
\end{equation*}
with $\Gamma_{\downarrow\uparrow} = aB^\mathrm{5}$, the relaxation rate from $\ket{\uparrow}$ to $\ket{\downarrow}$, and $\gamma = g_\mathrm{e}\mu_\mathrm{B}B / k_\mathrm{B}T$\,\cite{VasilisT1}. 
Here, $g_\mathrm{e}$ is the electron $g$-factor, $\mu_\mathrm{B}$ is the Bohr magneton and $T$ is the temperature. 
The relaxation rate pre-factor $a$ was found to be 0.08\,s$^\mathrm{-1}$T$^\mathrm{-5}$ derived from the effective-mass theory\,\cite{VasilisT1}. 
For \InZnN, the same dependence on $B$ is observed, however with an approximately four time smaller prefactor $a$. 
This result indicates the longitudinal spin relaxation mechanism is identical for both types of donors, with no degradation observed for In due to residual implantation damage. 
The difference in the prefactor is expected from the dependence of both the spin-orbit interaction and piezo-phonon coupling on the effective-mass wavefunction. 
A small dependence of \Tone\ on implantation fluence is also observed (see Appendix F), which is consistent with the dependence of \Tone\ on donor density observed in Ga \cite{VasilisT1}.

\subsection{Two-Laser Spectroscopy and Coherent Population Trapping (CPT)}

Two-laser spectroscopy can be further utilized to probe the spin properties of \InZnN. 
Fig.~\ref{fig:OPnCPT}(a) shows the results of resonant one-laser and two-laser spectroscopy performed at \SI{7}{\tesla} (Voigt geometry, $\vec{B} \perp \vec{c}$) and \SI{2}{\kelvin} for an nominal implantation fluence of 10$^{9}$\;cm$^{-2}$.
The PL dependence on $B$ and a discussion of the $g$-factors can be found in Appendix G. 
When one laser (denoted as S) scans the \InZnN X-related transitions, four transitions can be resolved, corresponding to transitions between the two D$^0$ electron spin states and two D$^0$X hole spin states (Fig.~\ref{fig:OPnCPT}(a)). 
The overall PL intensity, however, is much lower than what is observed for measurements without magnetic field. 
When a second laser (denoted as P) is resonant on the $\ket{\downarrow}$ to $\ket{\downarrow\uparrow\Uparrow}$ transition, an enhancement in signal is seen when the scanning laser scans the transitions connected to $\ket{\uparrow}$. 
Without the pump laser (one-laser measurement), population is transferred from one spin state of the ground state to the other, where it is trapped and cannot be re-excited. 
When the second laser is placed resonantly on the opposite spin state, population is transferred back to the spin state that the scanning laser can excite. 

The two-laser experiment performed here is often referred to as ``reverse spectral hole burning"\,\cite{YangSpectralHoleBurning}. 
The observed antihole can be used to determine the homogeneous linewidth of an optical transition in the presence of inhomogeneous broadening; the pump laser repopulates only the subensemble resonant with the the narrowband pump~\cite{YangSpectralHoleBurning}. 
A double Voigt fit to the antihole linewidth reveals a linewidth of 8\;GHz at \SI{2}{\kelvin}, surprisingly similar to the PLE linewidth at \SI{0}{\tesla} (see Fig.~\ref{fig:linewidth}~(a)). 
For the reported lifetime of \SI{1350}{\pico\second} for \InZnN X\,\cite{WagnerD0X}, the corresponding lifetime-limited linewidth is \SI{120}{\mega\hertz}. 
Additionally, \InZnN\ exhibits a strong hyperfine interaction of the bound electron with the spin-9/2 In nucleus, which splits the electron spin into ten levels spaced by \SI{50}{\mega\hertz}\,\cite{BlockInODMR, GonzalezInODMR}. 
The expected linewidth of the reverse spectral hole is \SI{550}{\mega\hertz} due to the combined lifetime-limited linewidth and the hyperfine interaction. 
Thus, the 8 GHz antihole indicates additional broadening mechanisms that will be the subject of future investigations.

In addition to the antihole peak, a narrow \SI{1}{\giga\hertz} dip can be observed on two-photon resonance (Fig.~\ref{fig:OPnCPT}(a) inset). 
The presence of the dip is the signature of coherent population trapping and the establishment of a ground state spin coherence~\cite{MariaZnO, FuCPT, SantoriCPT, GrayCPT, XuCPT}. Fig.~\ref{fig:OPnCPT}(b) displays higher resolution spectra for the CPT dip seen in Fig.~\ref{fig:OPnCPT}(a) for different excitation powers for the scanning and pump lasers. 
Also included is a CPT spectrum for in-grown Al$_\mathrm{Zn}^\mathrm{0}$ for reference. 
It is immediately evident that the CPT dip of \InZnN\ is much broader than the \SI{60}{\mega\hertz} \AlZnN\ dip. 
Moreover, for lower combined powers of the scanning and pump lasers, the overall shape of the In CPT dip resembles an inverted top hat, rather than the Lorentzian or Gaussian lineshape expected for a single homogeneously or inhomogeneously broadened transition. 
We attribute these unique features to the hyperfine interaction of the In donor electron with the spin-9/2 In nucleus. 
At high fields, in which the electron spin number is a good quantum number, two-photon Raman transitions correspond to dips occurring between states of the same nuclear spin. 
Thus, 10 dips separated by twice the hyperfine interaction are expected (see energy diagram in Appendix H). 
For Al$_\mathrm{Zn}^\mathrm{0}$ one expects 6 dips separated by  \SI{1.5}{\mega\hertz}\,\cite{OrlinskiiAlHF}. 
The In data shown in Fig.~\ref{fig:OPnCPT}~(b) are fit assuming ten Lorentzian dips spaced by \SI{100}{\mega\hertz}. In the lowest power In CPT spectrum, the best fit corresponds to a dip linewidth of 85\,MHz; however, reasonable fits can be obtained for linewidths ranging from 50 to 170\,MHz. 
Measurements of laser frequency drift and repeatability indicate the lack of resolved dips may be instrumentation limited at this time. 
However the flat CPT bottom confirms the expected 100 MHz hyperfine interaction and a potential path toward optical nuclear spin readout.

\section{Conclusion and Outlook}
In summary, we demonstrate that ion implantation and annealing can be used to form \InZnN\ ensembles with promising optical and spin properties for quantum information applications. 
The implanted \InZnN\ exhibit optical linewidths less than \SI{10}{\giga\hertz} for the \InZnN X transition, comparable with those of in-grown \InZnN. 
 $T_\mathrm{1}$ exceeds previously reported values for in-grown \GaZnN\,\cite{VasilisT1}, indicating that residual implantation damage has a negligible influence on longitudinal spin relaxation. 
Notably, the dominant longitudinal spin relaxation mechanism for \InZnN\ is the same process as what was reported for \GaZnN\,\cite{VasilisT1}, but with a distinctively lower overall relaxation rate. Using two-laser resonant excitation, we demonstrate that a coherent superposition of the ground states of \InZnN\ can be created via CPT. 
Power-dependent CPT measurements indicate that, for low laser powers, the CPT lineshape for \InZnN\ is determined by the hyperfine interaction between the donor bound electron and the spin-9/2 In nucleus. 
Thus, it may be possible in the future to access the nuclear spin degrees of freedom of implanted \InZnN\ optically. 
These results demonstrate that ZnO is a promising host material platform for low-damage creation of donor qubits via ion implantation and subsequent annealing, and thus the deterministic fabrication of donor spin qubits with optical access. 

\section{Acknowledgements}
The authors thank Yusuke Kozuka for the bulk ZnO substrates used for implantation, David Peterson for laser marking the ZnO substrate, and Nicholas S. Yama for assistance in sample processing. 
%This material is based upon work supported by the Army Research Office MURI Grant on Ab Initio Solid-State Quantum Materials: Design, Production and Characterization at the Atomic Scale (18057522) and the National Science Foundation under Grant 2212017. 
This material is based upon work primarily supported by the U.S. Department of Energy, Office of Science, Office of Basic Energy Sciences under Award No. DE-SC0020378, and partially supported by Army Research Office MURI Grant on Ab Initio Solid-State Quantum Materials: Design, Production and Characterization at the Atomic Scale (18057522) and the National Science Foundation under Grant 2212017.
This work was performed, in part, at the Center for Integrated Nanotechnologies, an Office of Science User Facility operated for the U.S. Department of Energy (DOE) Office of Science. 
Sandia National Laboratories is a multimission laboratory managed and operated by National Technology and Engineering Solutions of Sandia, LLC, a wholly owned subsidiary of Honeywell International, Inc., for the U.S. DOE's National Nuclear Security Administration under contract DE-NA-0003525. 
The views expressed in the article do not necessarily represent the views of the U.S. DOE or the United States Government.

\newpage

\appendix 

\section{Sample Layout}

\begin{figure}[htb]
	\centering
	\includegraphics[width=0.5\linewidth]{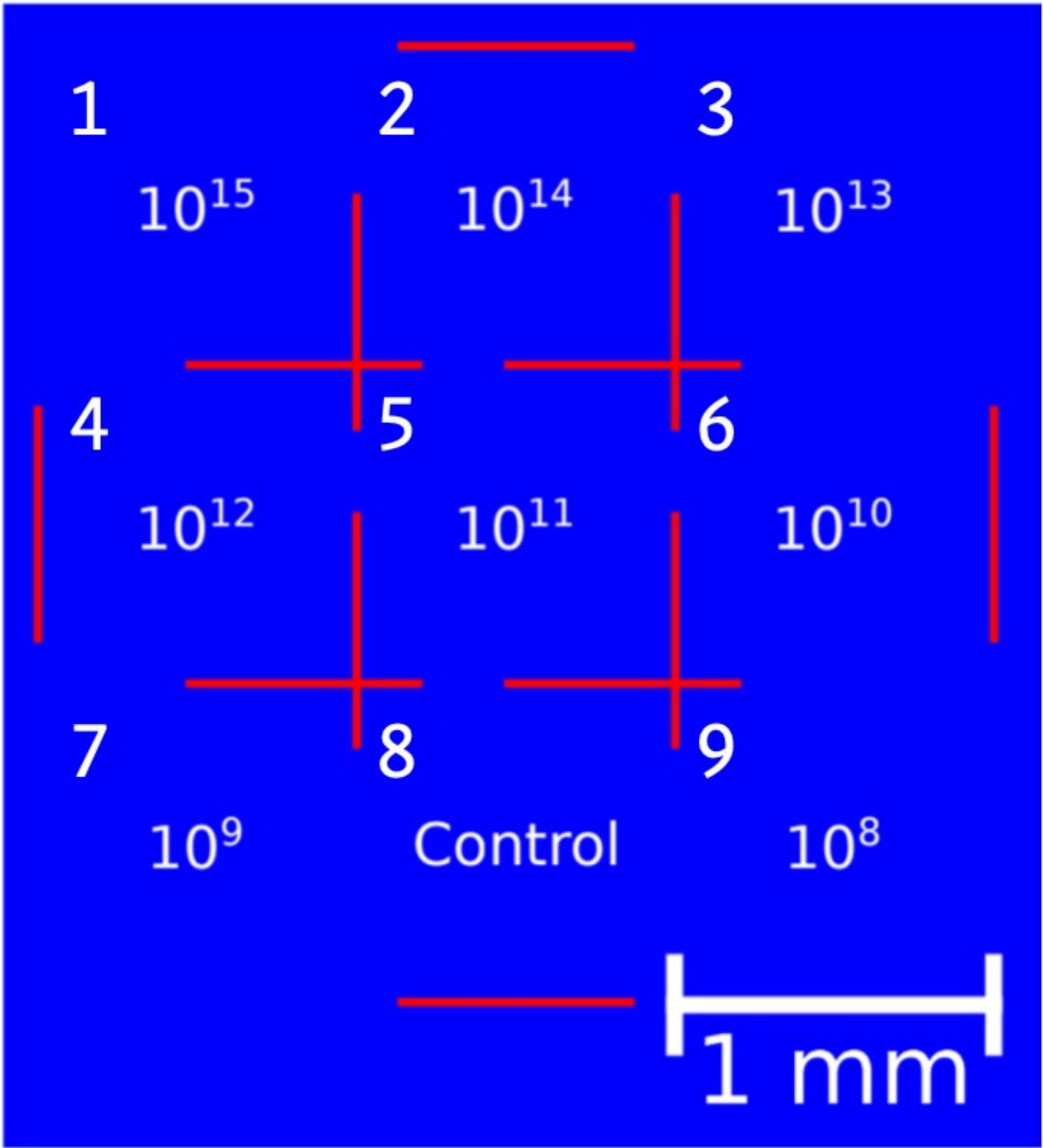}
	\caption{Layout of the grid.}
	\label{fig:layout}
\end{figure}

A grid pattern defining nine implantation areas is fabricated on the surface of the sample with a UV laser cutter (Protolaser U3) prior to implantation. 
Fig.~\ref{fig:layout} shows the layout. Tab.~\ref{tab:fluences} lists the nominal implantation fluences used for each region marked in Fig.~\ref{fig:layout}.

\begin{table}[!htb]
	\centering
	\begin{tabular}{ccc} 
		\hline
		\addlinespace[0.1em]
		Region & Nominal fluence (cm$^\mathrm{-2}$)& Measured fluence (cm$^\mathrm{-2}$)\\
		\addlinespace[0.1em]
		\hline
		\addlinespace[0.1em]
		1 & 10$^{15}$ & / \\
		2 & 10$^{14}$ & / \\
		3 & 10$^{13}$ & 1.1~$\times$~10$^{14}$ \\
		4 & 10$^{12}$ & 5.0~$\times$~10$^{12}$ \\
		5 & 10$^{11}$ & /\\
		6 & 10$^{10}$ & 1.7~$\times$~10$^{11}$ \\
		7 & 10$^{9}$ & /  \\
		8 & 0 & 1.3~$\times$~10$^{11}$\\
		9 & 10$^{8}$ & / \\
	\end{tabular}
	\caption{Summary of nominal implantation fluences and measured implantation fluences from SIMS for the different regions marked in Fig.~\ref{fig:layout}. ``/'' denotes not measured.}
	\label{tab:fluences}
\end{table}

\section{Photoluminescence Spectra with implantation fluence}  
\begin{figure}[!htb]
	\includegraphics[width=0.92\linewidth]{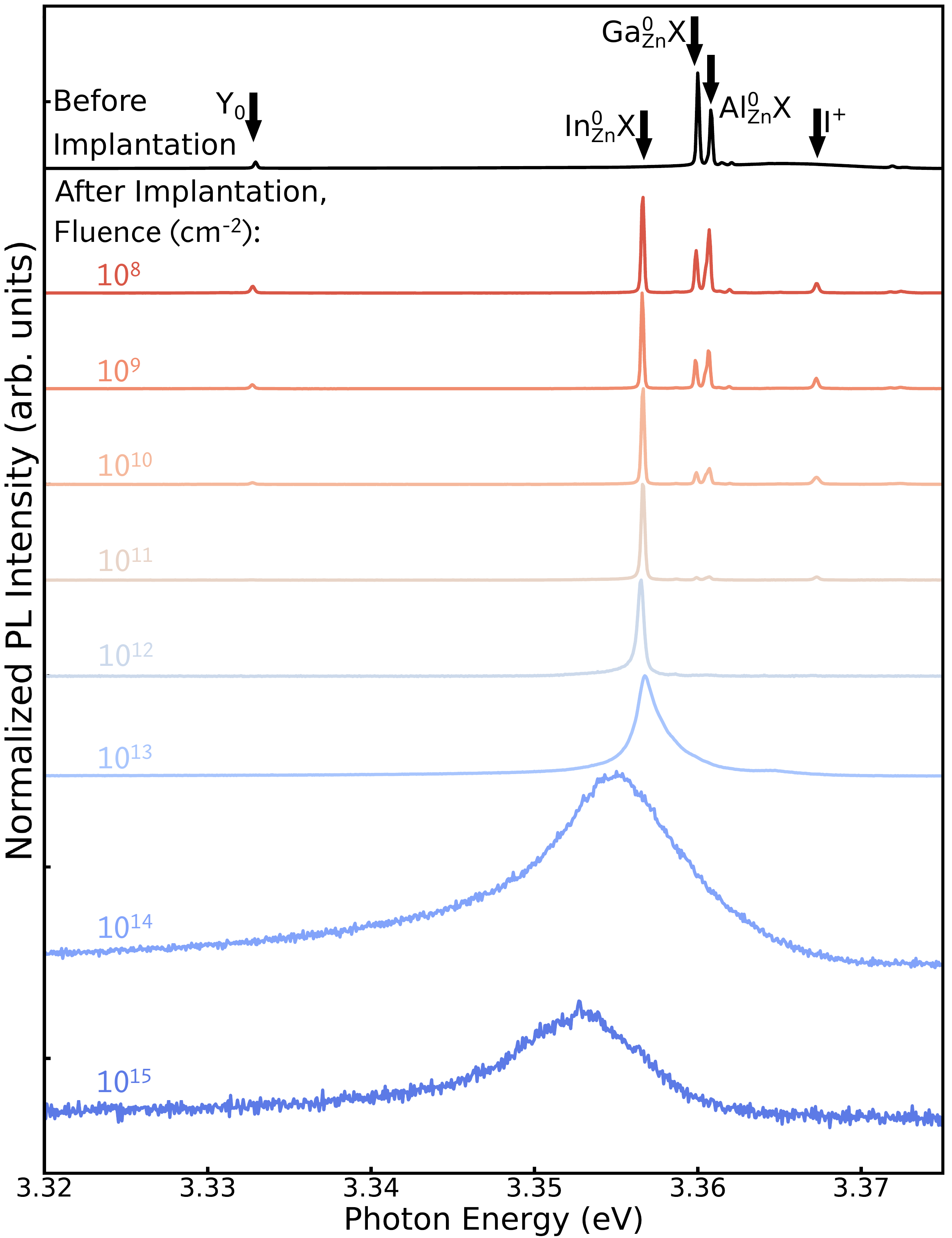}
	\caption{\label{fig:Sample_PL_overview} PL spectra for an as-grown ZnO single crystal and for the crystal after In implantation of different fluences. The maximum intensity of each background-subtracted spectrum is normalized unity. $T=7.5$\;K, $\lambda_{\text{exc}}=360$\;nm, $I_{\text{exc}}=800$\;nW. }
\end{figure}

Fig.~\ref{fig:Sample_PL_overview} shows normalized PL spectra for an as-grown ZnO single crystal and for different In implantation regions after annealing. 
Each spectrum corresponds to a region in Fig.~\ref{fig:layout}. The significant broadening of the \InZnN X line at higher fluences is evident, and only the \InZnN X transition can be observed for implantation fluences exceeding 10$^{11}$\;cm$^{-2}$.

\section{New donor lines after implantation}

\begin{figure}[h]
	\centering
	\includegraphics[width=1\linewidth]{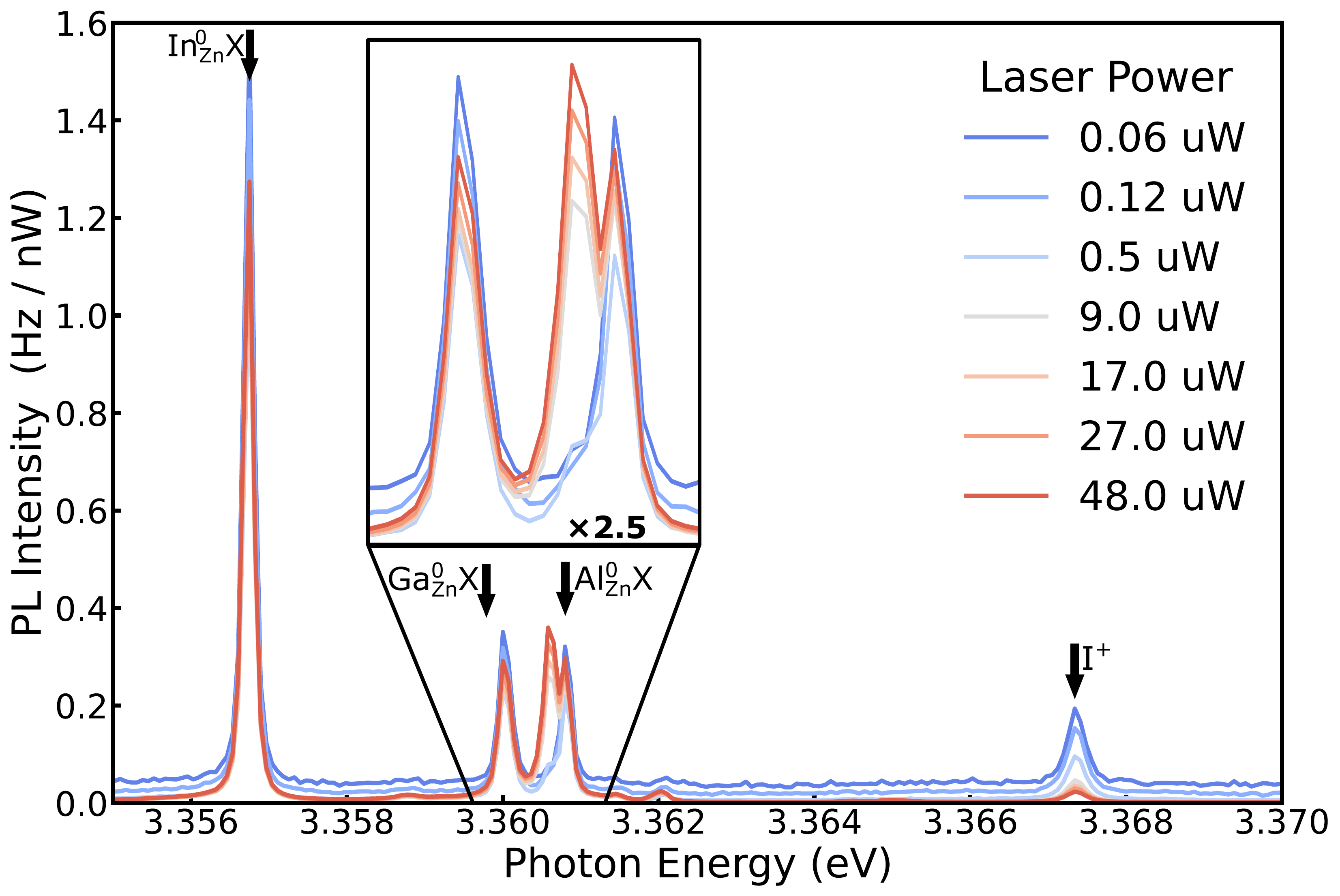}
	\caption{\label{fig:AboveBand_PowerSeries}PL spectra per laser power versus photon energy for 10$^{9}$\;cm$^{-2}$ implantation for different laser powers. $T=10.5$\;K, $\lambda_{\text{exc}}=360$\;nm.
	}
\end{figure}

Fig.~\ref{fig:AboveBand_PowerSeries} shows PL intensity versus excitation power for an implantation fluence of 10$^9$\;cm$^{-2}$. 
At powers above \SI{9}{\micro\watt}, a low energy shoulder on the Al peak ($\sim$\SI{3.3606}{\electronvolt}) grows with increasing laser power. 
This shoulder is not observed in nonimplanted samples. The origin of this new excitonic feature is not known, however it has a similar energy to the I$_7$ line related to carbon impurities that was observed in prior work \cite{Mohammadbeigi2014CRD}. If this peak is carbon related, it is unclear how carbon was introduced during the In implantation and annealing process.

\newpage 

\begin{figure}[!htb]
	\centering
	\includegraphics[width=1\linewidth]{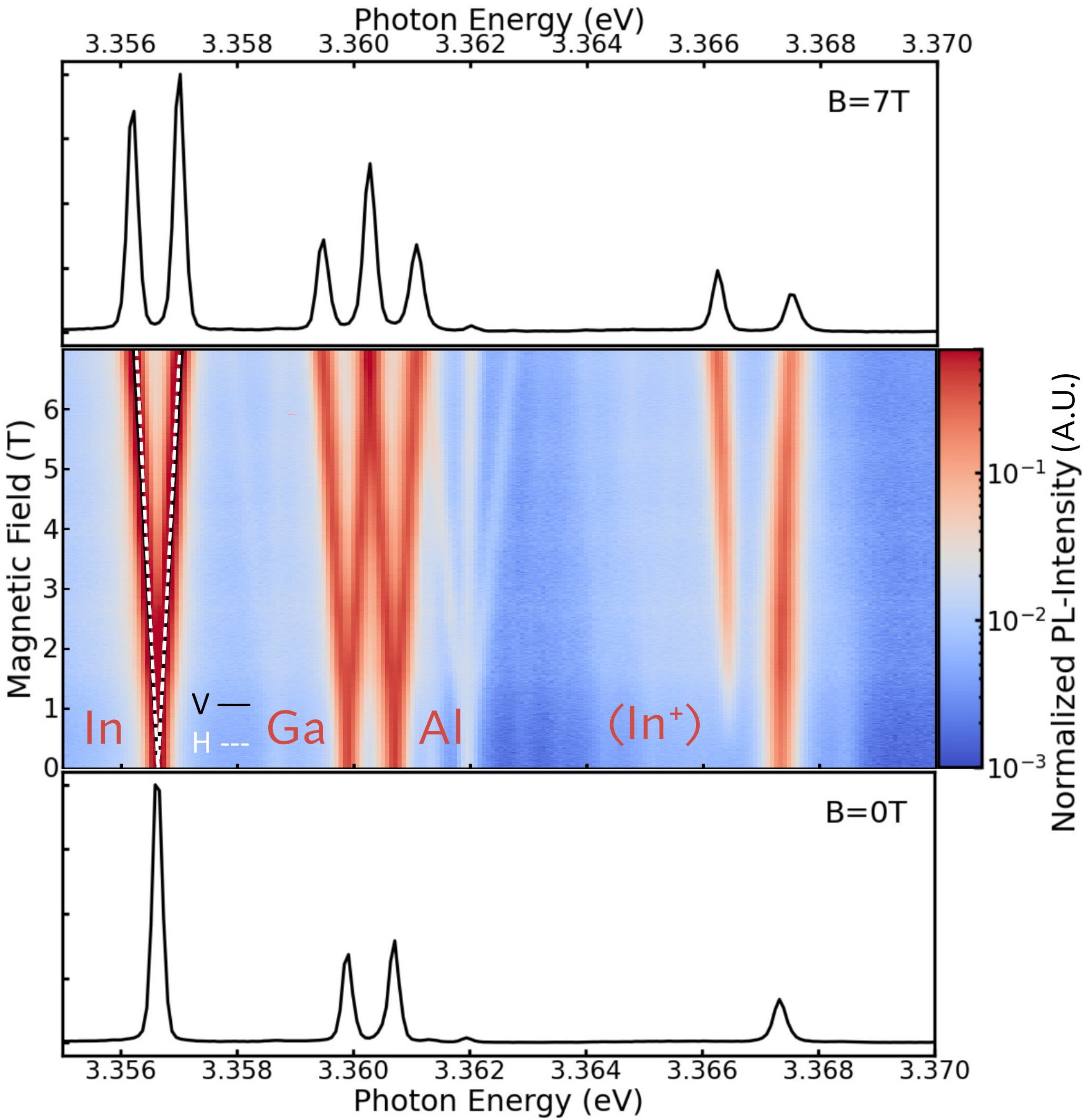}
	\caption{The bottom panel shows the above-band PL spectrum of the In-implanted ZnO crystal, $B=0$\,T. The middle panel shows a heatmap of the above-band PL at $B=0\;\text{to}\; 7$\,T. The dashed white lines and solid black lines denote the expected transitions based on the basis of the g factors of the electron and the hole determined in Appendix G. The top panel shows the above-band PL at $B=7$\,T. H, horizontal; V, vertical.}
	\label{fig:magnetoPL}
\end{figure}

The magneto-PL spectra at external magnetic fields in the Voigt geometry ($\vec{B} \perp \vec{c}$) from $B=0$\;T to $B=7$\;T in Fig.\ref{fig:magnetoPL} show the Zeeman splittings of transitions In$_{\text{Zn}}^{0}$X, Ga$_{\text{Zn}}^{0}$X and Al$_{\text{Zn}}^{0}$X. 
In addition, the appearance of a transition at nonzero field near 3.3664 eV suggests a forbidden transition at zero field is allowed at nonzero field. 
Both this new transition at 3.3664 eV and the I$^\mathrm{+}$ line at 3.3673 eV do not split at nonzero magnetic field. 
These observations are consistent with transitions that originate from excitons bound to ionized In donors In$_\mathrm{Zn}^\mathrm{+}$X. \cite{RodinaMagnetoPL}. 

\section{Relationship between inidum-galium PL ratio and PLE intensity}
\begin{figure}[!htb]
	\centering
	\includegraphics[width=1\linewidth]{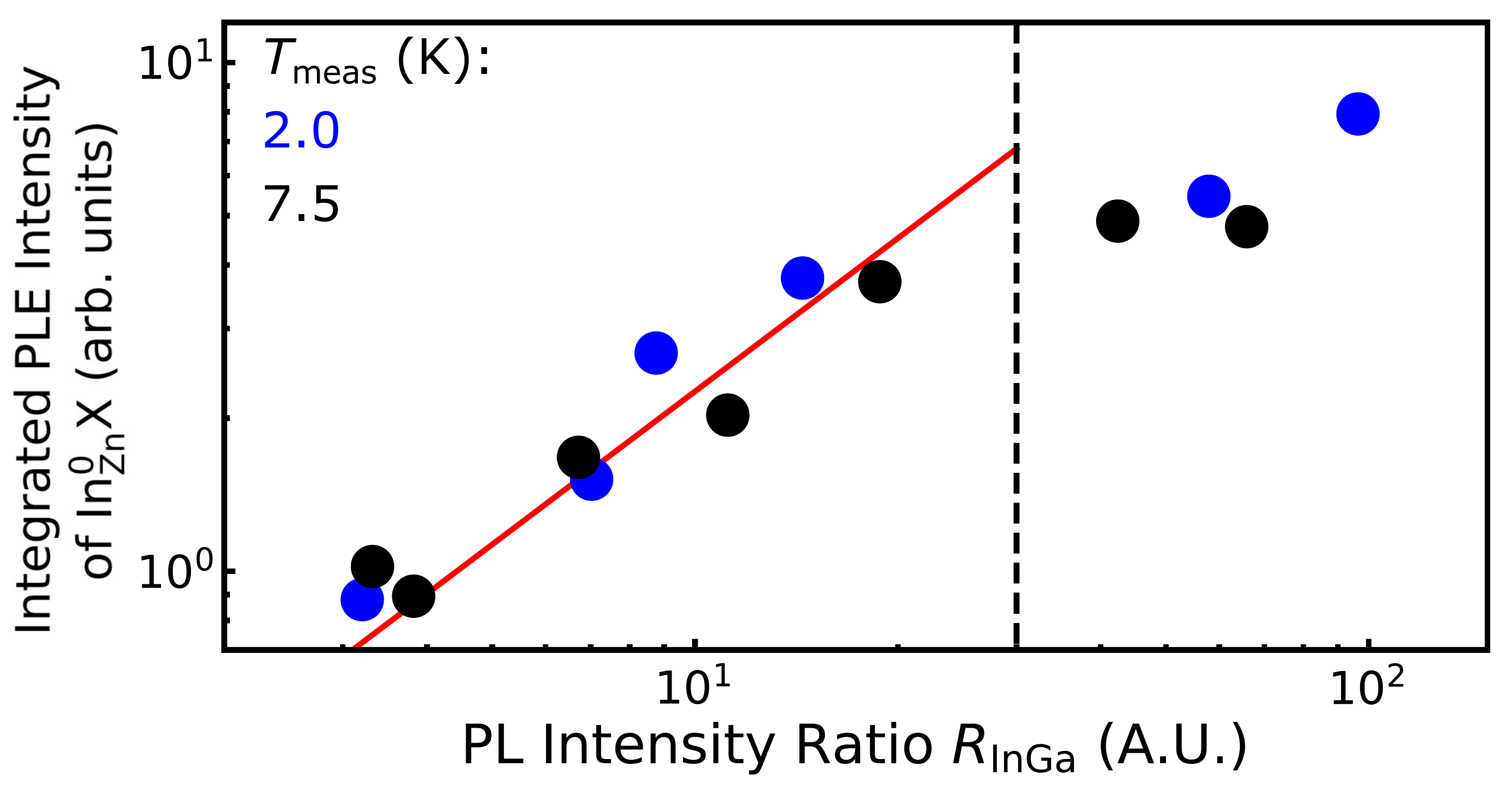}
	\caption{PLE intensity of \InZnN X as a function of the PL intensity ratio between the \InZnN X and \GaZnN X lines at \SI{2}{\kelvin} (blue dots) and \SI{7.5}{\kelvin} (black dots). 
		The red line depicts a linearly fitted curve.}
	\label{fig:PLEvsRInGa}
\end{figure}

We observe a linear relationship between the \InZnN X PLE intensity $A$ and the PL intensity ratio $R_{\text{InGa}}$ for low implantation fluences ($R_{\text{InGa}} < 30$). 
As shown in Fig.~\ref{fig:PLEvsRInGa}, by fitting a linear curve on 2\;K and 7.5\;K data, we estimate that $A = (0.23\pm 0.01) R_{\text{InGa}}$ for $R_{\text{InGa}} < 30$. 
$R_{\text{InGa}} < 30$ corresponds to nominal implantation fluences lower than \SI{5e+10}{\per\centi\metre\squared}.

\section{Photoluminescence Excitation Spectroscopy}

\begin{figure}[htb]
	\centering
	\includegraphics[width=0.45\textwidth]{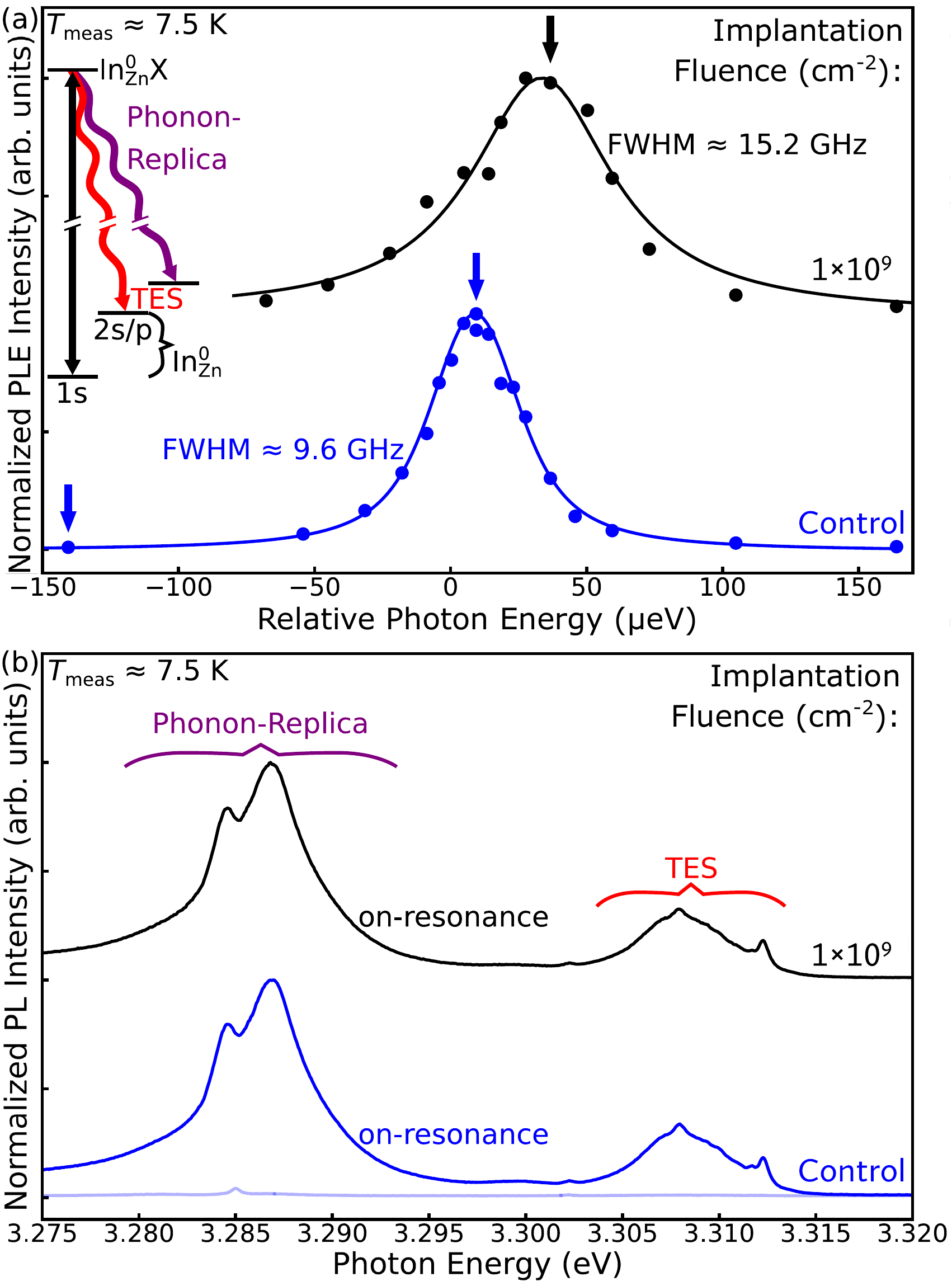}
	\caption{\label{fig:PLE}(a) PLE spectra integrating the TES and phonon replica signals (inset) for 10$^{9}$\;cm$^{-2}$ implantation and the control region. FWHMs of PLEs are obtained by fitted Voigt profiles.
		(b) Spectra of TES and phonon replica on- and off-resonance from \InZnN X taken at excitation wavelengths marked by arrows in (a). The maximum intensity of each background-subtracted spectrum is normalized to unity. $T=7.5$\;K. }
\end{figure}

Fig.~\ref{fig:PLE}(a) shows PLE spectra for implanted In$_\mathrm{Zn}^\mathrm{0}$X.
The PLE intensity is plotted by integrating the total signal of TES and phonon-replicas. Fig.~\ref{fig:PLE}~(b) displays the corresponding spectra for the laser on resonance and off resonance with the In$_\mathrm{Zn}^\mathrm{0}$X optical transition. 
The two-electron satellite transitions can only be observed for neutral donors, and would not be observed for \InZnP.

\section{Longitudinal spin relaxation time}

\begin{figure}[htb]
	\centering
	\includegraphics[width=0.9\linewidth]{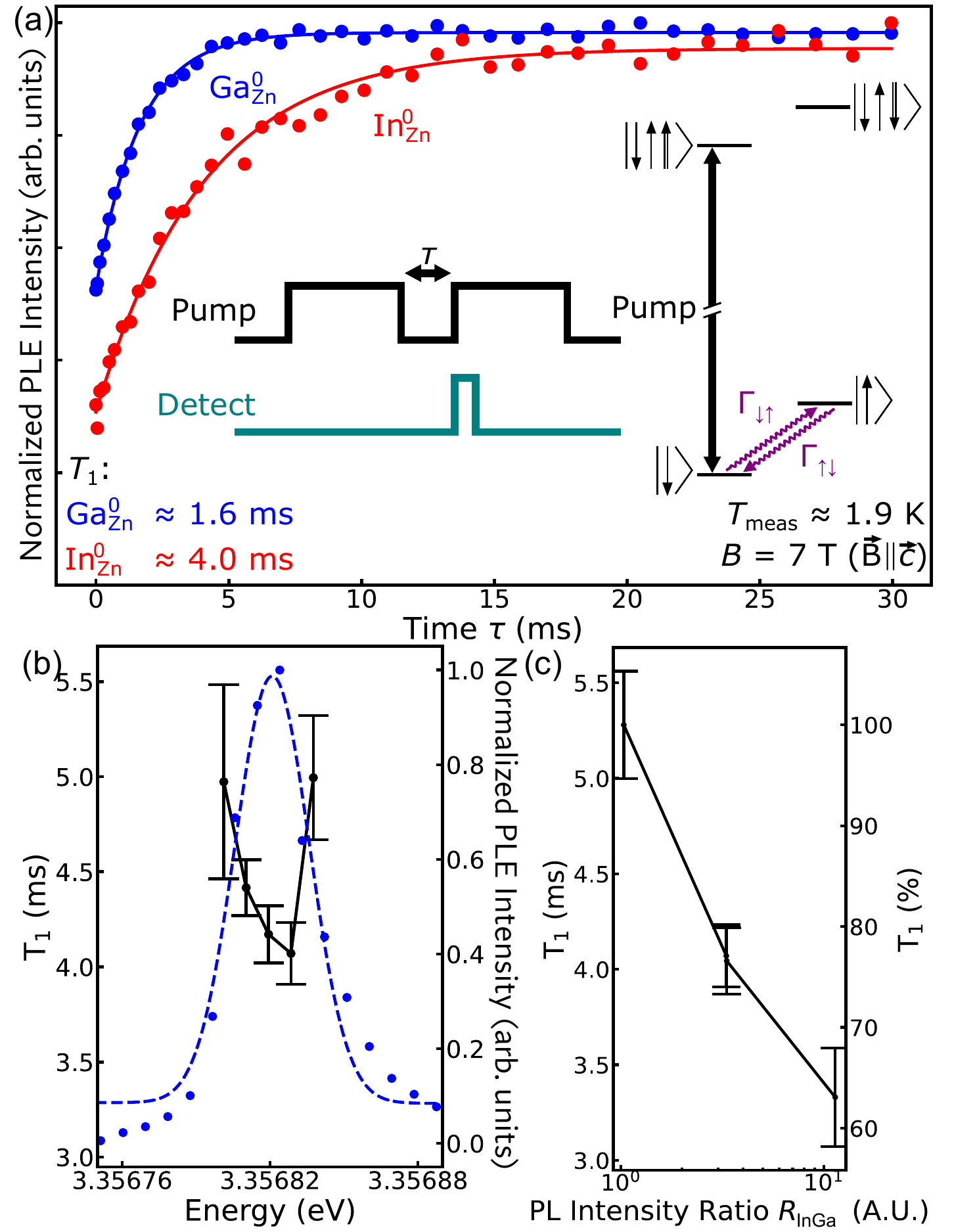}
	\caption{\label{fig:T1spec} 
		(a) Population recovery of state $\ket{\downarrow}$ from relaxation of the optically pumped state $\ket{\uparrow}$ as a function of delay time $\tau$ between each pump pulse for implanted In (10$^{9}$\;cm$^{-2}$) and in-grown Ga. The insets show the pump and detect sequence and the energy level diagram. $T_\mathrm{1}$ equals the decay constant of the fitted exponential profiles. 
		(b) Longitudinal spin relaxation time $T_1$ of \InZnN\ as a function of excitation photon energy (black points, left y-axis) and PLE spectrum (blue points, right y-axis) for a fluence of ~$10^9 \text{ cm}^{-2}$. Pump pulse on resonant with $\ket{\downarrow}\Leftrightarrow\ket{\downarrow\uparrow\Uparrow}$. 
		(c) $T_1$ as a function of the PL intensity ratio between \InZnN\ and \GaZnN\ denoted $R_{\text{InGa}}$.
		Different ratios correspond to different implantation fluences (see main text and Tab.~\ref{tab:fluences}).
		Error bars are one standard deviation of the exponential fit of $T_1$ curves in (a). Faraday geometry ($\vec{B}\parallel\vec{c}$), $B=7$\;T and $T=1.9$\;K. 
	}
\end{figure}

The longitudinal spin relaxation time $T_1$ of of {\it in situ}-doped \GaZnN\ donor ensembles was thoroughly investigated in~\cite{XiayuDirectT1, VasilisT1}. 
$T_1$ is determined by measuring the population recovery of spin state $\ket{\downarrow}$ after optical pumping of the population into the spin state $\ket{\uparrow}$ using resonant excitation for different delay times $\tau$ after the optical pumping pulse
(see Fig.~\ref{fig:T1spec}(a)).
The measurement scheme and energy level diagram are shown as insets in Fig.~\ref{fig:T1spec}(a). 

In prior work~\cite{VasilisT1}, we demonstrated an excitation-energy dependence that was attributed to varying effective donor density at a given resonant (or near-resonant) excitation energy.
In this work, we are able to directly investigate the donor density dependence of $T_1$ by probing locations of different implanted In fluences. 
In Fig.~\ref{fig:T1spec}(b) we show that for a given In fluence of ~$10^9\;\text{cm}^{-2}$, $T_1$ varies with excitation energy, the same as what was observed from Ga donors in previous studies. 
A shorter $T_1$ is observed when probed on resonance with the ensemble implanted In donors, indicating that a higher density of probed subensemble shortens the measured $T_1$.
Figure~\ref{fig:T1spec}(c) depicts the dependence of $T_1$ on the PL intensity ratio between \InZnN\ and \GaZnN, i.e. a proxy for implantation fluence. 
At higher implantation fluences, $T_1$ becomes shorter, which agrees qualitatively with our explanation for the effective donor density dependence of $T_1$.

\section{g-factors of neutral donor-bound excitons}

\begin{figure}[!htb]
	\centering
	\includegraphics[width=1\linewidth]{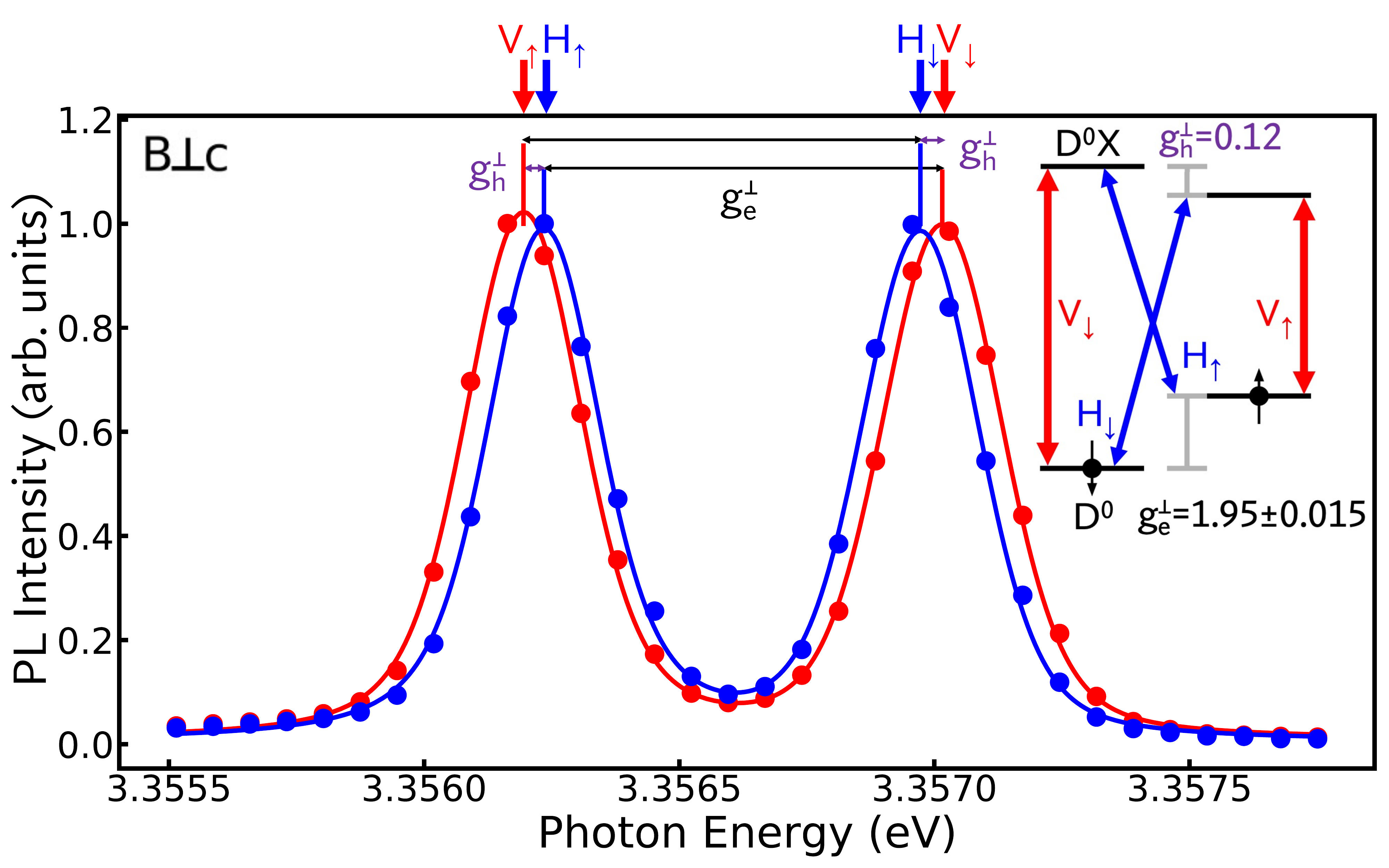}
	\caption{Magneto-PLs taken with different polarization optics show the energy separation used to find the g-factors of the electron and the hole using $\Delta  E=g\mu_{\text{B}}B$.
		Inset: Energy level diagram at $\vec{B} \perp \vec{c}$, $B>0$. The measured lower bound of the hole g-factor is 0.12 and the corresponding electron g-factor is 1.95.}
	\label{fig:gfactor}
\end{figure}

The \DoX\ Zeeman energy is determined by the unpaired hole spin. 
At nonzero external magnetic field perpendicular to the crystal axis (Voigt geometry $\vec{B} \perp \vec{c}$), the energy level configuration of the four possible donor bound exciton transitions and each of their polarization are labeled in the inset in Fig. \ref{fig:gfactor}. 
The splitting of the hole is small compared to that of the electron. 
At the experimentally applied fields, the separation between pairs of transitions V$_{\downarrow}$ and H$_{\downarrow}$, V$_{\uparrow}$ and H$_{\uparrow}$ cannot be resolved. 
As shown in Fig.~\ref{fig:gfactor}, by collecting vertical and horizontal polarizations, one can observe a shift in the excitonic emission.
This shift represents a lower bound of the g-factor of the hole, $g_{h}^{\perp}=0.12$.
The g-factor of the electron is 1.95, which are in good agreement with Ref.~\cite{RodinaMagnetoPL}. 

\section{Coherent Population Trapping}

\begin{figure}[!htb]
	\centering
	\includegraphics[width=1\linewidth]{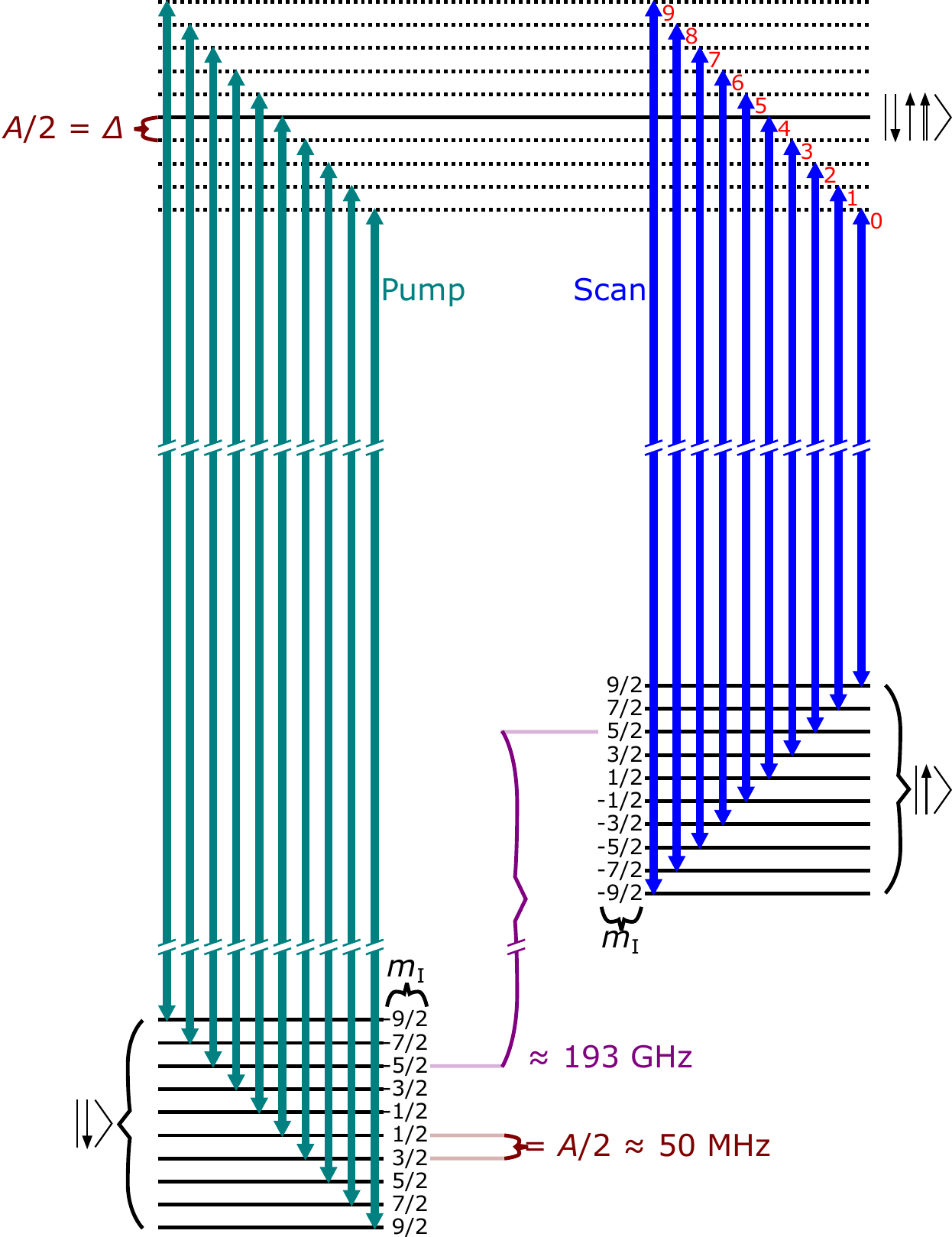}
	\caption{\label{fig:CPT_Scheme} Ten expected two photon resonances of CPT due to the hyperfine interaction between the valence electron of In$_\mathrm{Zn}^\mathrm{0}$ and In nuclei of the same m$_\mathrm{I}$.}
\end{figure}

Fig.~\ref{fig:CPT_Scheme} shows a schematic illustration of the two-photon transitions in the \InZnN/\InZnN X system due to the hyperfine interaction between In nuclei (all stable In isotopes have $I$~=~9/2) and the valence electron of \InZnN\cite{BlockInODMR, GonzalezInODMR}.
It is assumed that the two-photon resonance necessary for CPT can occur only between ground state levels with the same nuclear spin quantum number (m$_\mathrm{I}$), i.e., nuclear spin flips are forbidden. 
For a constant wavelength of the pump laser, this leads to different resonance conditions depending on m$_\mathrm{I}$, i.e., different resonant wavelength for the scanning laser. We expect ten resonances equally spaced by $A$. 
CPT can occur at different detunings from the single level representing $\ket{\downarrow \uparrow \Uparrow}$. The hyperfine splitting of the excited state is expected to be much smaller than that of the ground state due to the spin-singlet nature of the electrons and the p-orbital nature of the hole.

\newpage

\bibliography{main.bib}

\end{document}